\documentclass[11pt,a4paper]{article}

\usepackage{amsfonts}
\usepackage{graphicx}
\usepackage{float}

\title{\textbf{Higgs phase in a gauge $\mathbf{U}(1)$ non-linear $\mathbf{CP}^1$-model.\\
 Two species of BPS vortices and their zero modes.}}

\author{A. Alonso Izquierdo$^{(a)}$ and J. Mateos Guilarte$^{(b)}$
\\ {\normalsize {\it $^{(a)}$ Departamento de Matematica
Aplicada}, {\it University of Salamanca, SPAIN}}\\  {\normalsize {\it $^{(b)}$ Departamento de Fisica
Fundamental}, {\it University of Salamanca, SPAIN}}}

\date{}

\begin{document}

\maketitle

\begin{abstract}
In this paper zero modes of fluctuation are dissected around the two species of BPS vortices existing in the critical Higgs phase, where the scalar and vector meson masses are equal, of a gauged $\mathbb{U}(1)$
nonlinear $\mathbb{CP}^1$-model. If $2\pi n$, $n\in \mathbb{Z}$, is the quantized magnetic flux of the two species of BPS vortex solutions, $2n$ linearly independent vortex zero modes for each species are found and described. The existence of two species of moduli spaces of dimension $2n$ of these stringy topological defects is thus locally shown.
\end{abstract}


\section{Introduction}

$\mathbb{CP}^N$ models in two-dimensional Euclidean space-time were introduced by Golo and Perelomov, see \cite{Golo}, with the goal of discussing self-dual $2D$-instantons in a complex/Kh$\ddot{\rm a}$ler framework. These systems in Minkowski $\mathbb{R}^{1,3}$ space-time are a variant of the $\mathbb{O}(N)$ non-linear sigma model, which is constructed in turn as a restriction of the linear sigma model of Gell-Mann and Levy \cite{Gell-Mann} to the dynamics of their Goldstone bosons. Several generalizations of the non-linear $\mathbb{O}(N)$-sigma model appeared as low energy effective theories in hadronic physics related to current algebras before the surge of non-Abelian gauge theories in particle physics. In this work we shall focus in a model belonging to the family of gauged non-linear Sigma models, see e.g. \cite{Percacci}, which might be treated also in interaction with gravity, see \cite{Percacci1}, granting to this kind of systems a r$\hat{\rm o}$le in Cosmology.

In particular, we shall concentrate in the simplest model studied in Reference \cite{Alonso} where we added to the action of the gauge $\mathbb{CP}^1$-model a potential energy density depending only on the fields and guaranteeing that both the scalar and vector particles are massive. Moreover, we showed that in this model there are two species of BPS vortices living respectively in the south and north charts. Because our vortex solutions have cylindrical symmetry they can be interpreted as cosmic strings conveying by themselves cosmological implications if the model is defined in curved space-time, see {\it inter alia} the monographic textbook \cite{Vilenkin} and References quoted therein.

Besides of letting the system to live in the Higgs phase, our choice of the potential energy density is guided by the existence of a self-dual structure: first-order static field equations together with a topological bound. This structure arises at the critical point where the scalar and vector particle masses are equal and it is the requirement to build generalizations of these gauge non-linear sigma models with extended supersymmetry. In fact, in the disguise of $2D$ instantons Nitta and Vinci discovered in \cite{nivi} the same topological defects in the framework of two-dimensional ${\cal N}=(2,2)$ SUSY sigma model that we described within a purely bosonic context in \cite{Alonso}. The promotion of the field theoretical model at the stake to a ${\cal N}=2$ supersymmetric status is possible through an standard procedure using the K$\ddot{\rm a}$hler structure of $\mathbb{CP}^1$.

Our goal in this essay, however, is to analyze thoroughly the BPS vortex zero modes of the two species. Despite of being in a Higgs phase
our model exhibits zero gap fluctuations around the BPS vortices. In Reference2 \cite{AlGarGuil1, AlGarGuil2} we developed this task in a fully detailed manner for the BPS vortices in the Abelian Higgs model. The information that we acquired in this work was the key to the computations achieved in \cite{AlGuilTor} of the one-loop BPS vortex string tension shifts in the AHM when these objects are immersed in the quantum world. We plan thus to establish the analogies and differences between the zero modes of fluctuation of the standard self-dual vortices
and those of the two species of BPS vortices in our gauged $\mathbb{CP}^1$ model. Because zero modes provide a bridge between the classical and quantum domains we hope that the results obtained here will be useful in future research about quantum properties of the two species of BPS $\mathbb{CP}^1$ vortices.

\section{A gauge $\mathbb{U}(1)$ massive non-linear $\mathbb{CP}^1$-sigma model with two self-dual vortex species}

We shall address a system of three scalar fields $\Phi=(\Phi_1,\Phi_2,\Phi_3)$ which take values on a $\mathbb{S}^2$-sphere target space:
 \[
 \Phi(x^\mu): \mathbb{R}^{1,3} \rightarrow  \mathbb{S}^2 \, \, \, \, \mbox{where} \, \, \, \,  \Phi_1^2(x^\mu)+\Phi_2^2(x^\mu)+\Phi_3^2(x^\mu)=\rho^2 \, \, \, \mbox{and} \, \, \,  x^\mu\equiv (x^0,x^1,x^2,x^3)\in \mathbb{R}^{1,3} \, .
 \]
We also denote $x^\mu\cdot x_\mu=g_{\mu\nu}x^\mu x^\nu$ and choose the metric tensor in $\mathbb{R}^{1,3}$ as: $g_{\mu\nu}={\rm diag}(1,-1,-1,-1)$. Stereographic projections respectively from the north and south poles of the sphere to the plane lead to the south and north charts that form a minimal atlas in $\mathbb{S}^2$ intersecting at the equator. In these two planes/charts we define the complex scalar fields
\[
\phi^S(x^\mu)=\rho\frac{\Phi_1(x^\mu)+i \Phi_2(x^\mu)}{\rho-\Phi_3(x^\mu)}\hspace{0.5cm} , \hspace{0.5cm}
\phi^N(x^\mu)=\rho\frac{\Phi_1(x^\mu)-i \Phi_2(x^\mu)}{\rho+\Phi_3(x^\mu)} \hspace{0.5cm}  , \hspace{0.5cm}  \rho>0 \, \, .
\]
distinguished by the superscripts $S$ and $N$. Use of complex coordinates characterizes the target $\mathbb{S}^2$-sphere as the complex manifold $\mathbb{CP}^1$, i.e., the complex projective line.
The information coming from one of these charts is translated into the other one via the transition function $\phi^S={\rho^2}/{\phi^{N*}}$ that allows the global definition of the $\mathbb{CP}^1$ by prescribing how the two charts are related. We remark that our choice of transition function reverses the orientation of the north chart with respect to that of the south chart. The reason for choosing this option is to deal later with scalar fields coupled to the gauge field with identical electric charges in both charts. The massless Lagrangian ${\cal L}_0$ describing the dynamics of the $\mathbb{CP}^1$ model written in terms of the south chart fields, reads:
\[
{\cal L}_0^S[\phi^S]=\frac{1}{2}\frac{4\,\rho^4}{(\rho^2+|\phi^S|^2)^2}\, \partial_\mu \phi^{S*} \cdot \partial^\mu \phi^S.
\]
The global $\mathbb{U}(1)$-symmetry with respect to changes in the field phase $\phi^S \rightarrow e^{i\chi}\phi^S$ may be promoted to local invariance following the standard procedure: a gauge field $A_\mu$ enters the system and supplements the local $\mathbb{U}(1)$ transformation $\phi^S \rightarrow e^{i\chi(x^\mu)}\phi^S$ with the gauge transformation $A_\mu\rightarrow A_\mu+\partial_\mu\chi$. A potential energy density yielding spontaneous breaking of the gauge symmetry in the Higgs mechanism mode will be also allowed. All this leads to the Lagrangian of a gauged  massive Abelian non-linear $\mathbb{CP}^1$ model with dynamics governed by the Lagrangian:
\begin{equation}
{\cal L}^S[\phi^S,A_\mu]=-\frac{1}{4} F_{\mu\nu} F^{\mu\nu} +\frac{1}{2}\frac{4\rho^4}{(\rho^2+|\phi^S|^2)^2} D_\mu \phi^{S*} D^\mu \phi^S-U_S(|\phi^S|^2)\label{lagphi}\, \, .
\end{equation}
The covariant derivatives and the electromagnetic tensor are defined as usual: $D_\mu\phi=\partial_\mu\phi-i A_\mu\phi$ and $F_{\mu\nu}=\partial_\mu A_\nu-\partial_\nu A_\mu$. In order to write the Lagrangian (\ref{lagphi}) in the north chart we need to know how the covariant derivatives and the potential energy density read there:
\[
D_\mu\phi^S=-\frac{\rho^2}{(\phi^{N*})^2}D_\mu\phi^{N*} \hspace{0.5cm}\mbox{and}\hspace{0.5cm} U_N(|\phi^N|^2)=U_S\left({\textstyle\frac{\rho^4}{|\phi^N|^2}}\right)\label{cambioderi} \, \, .
\]
Thus, in the other chart the Lagrangian becomes:
\begin{equation}
{\cal L}^N[\phi^N,A_\mu]=-\frac{1}{4} F_{\mu\nu} F^{\mu\nu} +\frac{1}{2}\frac{4\rho^4}{(\rho^2+|\phi^N|^2)^2} D_\mu \phi^{N*} D^\mu \phi^N-U_N(|\phi^N|^2)\label{lagpsi}
\end{equation}
a formula showing, together with (\ref{lagphi}), that this model is globally defined on $\mathbb{CP}^1$. For the sake of concision we shall write the Lagrangians (\ref{lagphi}) and (\ref{lagpsi}) in the unified form
\begin{equation}
{\cal L}[\phi,A_\mu]=-\frac{1}{4} F_{\mu\nu} F^{\mu\nu} +\frac{1}{2} g(|\phi|^2) D_\mu \phi^* D^\mu \phi+ U(|\phi|^2)\label{laggen}
\end{equation}
where $\phi$ stands for either $\phi^S$ or $\phi^N$ if either the south or the north charts are alternatively considered in the description of the model. In both charts the metric factor in the target manifold which appears
multiplying the kinetic terms respectively in (\ref{lagphi}) and in (\ref{lagpsi}) is written in the form:
\[
g(|\phi|^2) = \frac{4\rho^4}{(\rho^2 + |\phi|^2)^2} \, .
\]
Notice that the Abelian Higgs model Lagrangian follows the generic form (\ref{laggen}) when the metric factor $g(|\phi|^2)$ is the unity and there is only one chart in the non-compact target space $\mathbb{C}$. We recall that the celebrated Abrikosov-Nielsen-Olesen vortex filaments arise in the AHM as cylindrically symmetric solutions of the field equations grown from planar topological solitons endowed with a quantized magnetic flux. The ANO vortices are static configurations, henceforth their investigation requires the temporal gauge $A_0=0$. Axial symmetry is realized in two steps: first, one searches for field configurations living in the $x_3=0$ plane: $\phi=\phi(x_1,x_2)$ and $A_j=A_j(x_1,x_2)$, $j=1,2$, a restriction that is consistent only in the axial gauge $A_3=0$. Second, in the restricted second-order field equations one looks for solutions with this characteristics and appropriate boundary conditions, which force the quantization of the magnetic flux. Repeated infinitely in the third dimension the ANO magnetic flux tubes are obtained from these planar vortex solutions having finite energy per unit length. If the parameters of the AHM are such that the system lives at the critical point between Type I and Type II superconductivity, there exists a first-order PDE (Bogomolny) system of field equations such that the vortex solutions, usually called self-dual or BPS vortices, saturate a topological bound, i.e., their energy per length unit is proportional to the magnetic flux.

Closely following the developments in \cite{Alonso}, we summarize the investigation of BPS vortex solutions starting from the Lagrangian (\ref{laggen}), which covers the two species of BPS vortices in our gauged massive Abelian $\mathbb{CP}^1$ model.
From this Lagrangian one easily derives the energy per unit of length functional for static and planar configurations:
\begin{equation}
V[\phi,A_j] = \int_{\mathbb{R}^2} d^2 x \Big[\frac{1}{2} F_{12}^2 + \frac{1}{2} g(|\phi|^2) (D_j\phi)^* D_j\phi + U(|\phi|^2) \Big] \, \, .
\label{energyfunctional}
\end{equation}
We then focus on the set of all the static planar configurations with finite energy per unit length:
${\cal C}=\{(\phi,A):V[\phi,A]<+\infty\}$. Membership to ${\cal C}$ space demands the following asymptotic behavior:
\begin{equation}
\lim_{r\rightarrow \infty} \phi^* \phi=\vert v\vert^2, \hspace{0.3cm}\, \hspace{0.3cm} v \in {\cal M}=\{v\in \mathbb{C}: U(|v|^2)=0 \} \hspace{0.6cm}\mbox{and}\hspace{0.6cm} \lim_{r\rightarrow \infty} D_j\phi =0  \, \, \, , \, \, \, r=\sqrt{x_1^2  +x_2^2} \, ,\label{asintotico}
\end{equation}
where ${\cal M}$ is the set of zeroes of the non negative function $U(\vert \phi\vert^2)\geq 0$. Moreover, we shall restrict ourselves to a subclass of models governed by the Lagrangian (\ref{laggen}) for which the energy per unit length functional (\ref{energyfunctional}) admits a Bogomolny arrangement. It was proved in Reference
 \cite{Alonso} that $V[\phi,A]$ is the sum of positive (perfect squares) terms plus a (positive) topological quantity:
\begin{equation}
V[\phi,A]=\frac{1}{2}\int_{\mathbb{R}^2} \, d^2x \, \left\{\Big(F_{12}\pm\frac{1}{2}\Big[G(\phi^*\phi)-\alpha^2\Big]\Big)^2+g(\vert\phi\vert^2)\big\vert D_1\phi\pm i D_2 \phi\vert^2\right\}+\frac{\alpha^2}{2} \,\,\Big\vert \int_{\mathbb{R}^2} \, d^2x \, F_{12}\Big\vert \label{bogs} \, \, ,
\end{equation}
when the scalar field potential term $U(|\phi|^2)$ is of the form
\begin{equation}
U(\phi)=\frac{1}{8} \Big[ G(|\phi|^2)-\alpha^2 \Big]^2 \hspace{0.5cm} \mbox{where} \hspace{0.5cm} G(z)=\int_0^z dz \, g(z) \, \, ,\label{potential}
\end{equation}
having into account that $g(\vert\phi\vert)> 0$.
Notice that $G(0)=0$, $G'(z)=g(z)$ because of the definition in (\ref{potential}) and we choose $0<\alpha^2<G(\infty)$ in order to leave room for spontaneous symmetry breaking of the $\mathbb{U}(1)$ symmetry.
Solutions of the first-order PDE system
\begin{equation}
D_1\phi\pm i D_2 \phi =0 \hspace{0.5cm},\hspace{0.5cm} F_{12}\pm \frac{1}{2} \Big[ G(|\phi|^2) -\alpha^2 \Big]=0 \label{pde01}
\end{equation}
are thus minima of the energy per unit length and therefore also solve the second-order field equations. Moreover, the boundary conditions (\ref{asintotico}) ensure that the vector field $A_j \simeq_{r\to \infty} i\phi^*\partial_j\phi$ is asymptotically purely vorticial, and, via Stoke's theorem, that the total magnetic flux is a topological quantity given by the degree of the map from the circle at infinity $S^1_\infty$ to the manifold ${\cal M}$ provided by the asymptotic behaviour of the scalar field. BPS or self-dual vortices are static and $x_3$-independent field configurations that solve the PDE system (\ref{pde01}) subjected to
the boundary conditions (\ref{asintotico}).  Thus, the BPS vortex solutions saturate the Bogomolny bound $V[\phi^{BPS}, A^{BPS}]=\alpha^2 \pi n$, where $n$ is an integer, becoming the configurations of minimum energy per unit of length in each disconnected sector of the configuration space. The BPS magnetic flux lines are accordingly stable minima of the functional $V(\phi,A)$ which do not exhibit fluctuation modes of negative energy. Saturation of the Bogomolny bound, however, is compatible with flat directions in configuration space ${\cal C}$ of neutral equilibrium. In \cite{Alonso} we applied the index theorem unveiling the existence of $2n$ zero modes of fluctuations around the new species of vortices. This means that the quanta of magnetic flux are, at least locally, free of moving independently from each other through the moduli space of vortices.

Following the same strategy as in Reference \cite{AlGarGuil1} on the research for vortex zero modes in the AHM we shall first investigate the planar PDE system (\ref{pde01}) by using polar coordinates in the $x_1$-$x_2$ plane,
\begin{equation}
D_r \phi = \mp \frac{i}{r} D_\theta \phi \hspace{0.5cm},\hspace{0.5cm} \frac{1}{r} F_{r\theta} = \mp \frac{1}{2} \Big[G(|\phi|^2)-\alpha^2\Big] \label{pde02} \hspace{0.5cm},\hspace{0.5cm} \theta={\rm arctan}\frac{x_2}{x_1} \, \, ,
\end{equation}
i.e., we shall explore circularly symmetric, or, cylindrically symmetric seen in three dimensions, BPS vortices. The remnant gauge freedom allows us to set the radial gauge condition $A_r=0$ and we shall plug the ansatz
\begin{equation}
\phi(r,\theta)=f_n(r) \, e^{in\theta} \hspace{0.5cm} ,\hspace{0.5cm} r A_\theta(r) = n \, \beta_n(r) \hspace{0.5cm},\hspace{0.5cm} n \in \mathbb{Z} \label{ansatz06}
\end{equation}
into the PDE system (\ref{pde02}). We end with the non linear ODE system
\begin{equation}
\frac{d f_n}{d r} = \pm \frac{n}{r} f_n(r) [1-\beta_n(r)] \hspace{0.5cm},\hspace{0.5cm} \frac{n}{r} \frac{d \beta_n}{d r} = \frac{1}{2} \Big[ \alpha^2 - G[f_n^2(r)] \Big]
\label{pde03}
\end{equation}
determining the radial profiles $f_n$ and $\beta_n$ of the scalar and vector field components of the self-dual vortices. From the boundary conditions (\ref{asintotico}) we derive the asymptotic behavior of these radial profiles :
\begin{equation}
\lim_{r\rightarrow \infty} f_n(r)=f_\infty= [G^{-1}(\alpha^2)]^\frac{1}{2}  \hspace{0.5cm},\hspace{0.5cm} \lim_{r\rightarrow \infty} \beta_n(r)=1 \, \, \, , \label{radbc}
\end{equation}
where $G^{-1}$ denotes the inverse function to $G$. Because the vacuum orbit ${\cal M}=\{v: \,\vert v\vert^2=G^{-1}(\alpha^2)\}$ is a circle in the target manifold, the scalar field at infinity $\phi(\infty , \theta)=\sqrt{G^{-1}(\alpha^2)}\cdot e^{i n \theta}$ provides a map from the boundary of the plane at infinity to ${\cal M}$ which belongs to the $n$-th homotopy class of the first homotopy group of the circle: $\Pi_1(\mathbb{S}^1)=\mathbb{Z}$.  The winding number  $n$ of this map is also encoded in the fact that the magnetic flux is classically quantized:
\[
\Phi_{\rm magnetic} = \frac{1}{2\pi} \int_{\mathbb{R}^2} d^2x F_{12} =\frac{1}{2\pi}\lim_{r\to\infty}\oint_{\partial\mathbb{R}^2\equiv\mathbb{S}_\infty^1} r A_\theta(r)d\theta= n \, \, .
\]
We do not only require the asymptotic behavior (\ref{radbc}) to the vortex solutions of (\ref{pde03}), but we demand also regularity at the origin, which implies  $\beta_n(0)=0$ in (\ref{ansatz06}). But near the $(0,0,x_3)$ axis, the ODE system (\ref{pde03}) can be solved analytically because if $\beta_n(0)=0$ the equation for the radial profile becomes: $\frac{df_n}{dr} \approx \pm \frac{n}{r} f_n(r)$, which is solved by
\begin{equation}
f_n(r)= d_n r^n + o(r^{n+1}) \label{fseries} \,
\end{equation}
where $d_n$ is an integration constant.
We obtain next $\beta_n(r)$ near the origin by solving the other linearized equation in (\ref{pde03}):
\begin{equation}
\frac{n}{r} \frac{d \beta_n}{dr}= \frac{1}{2} \Big[ \alpha^2+ g(0)\, d_n^2 r^{2n} + o(r^{2n+1})\Big] \label{bseries0}
\end{equation}
To derive this linearization we used the radial profile (\ref{fseries}) near the origin and the power series expansion of the $G[f_n^2(r)]$ function around $r=0$: $G[f_n^2(r)]=g(0) d_n^2 r^{2n} + o(r^{2n+1})$ up to the $r^{2n+3}$-th order. Solving for $\beta_n(r)$ the linear ODE
(\ref{bseries0}) is an easy task:
\begin{equation}
\beta_n(r) = e_2 \,r^2 + e_{2n+2}\, r^{2n+2} + o(r^{2n+3}) \hspace{0.5cm} \mbox{where} \hspace{0.5cm} e_2= \frac{\alpha^2}{4n} \hspace{0.2cm},\hspace{0.2cm} e_{2n+2}= \frac{g'(0) \,d_n^2}{4n(n+1)} \label{bseries}
\end{equation}
The vortex radial profiles written in (\ref{fseries}) and (\ref{bseries}) clearly show that $f_n(0)=\beta_n(0)=0$, i.e., both the scalar and the vector vortex fields vanish at the vortex center $(x_1,x_2)=(0,0)$ as demanded.
The behavior in the intermediate region varies with the choice of potential $U(|\phi|^2)$, henceforth with the metric $g(|\phi|^2)$
of the target manifold. Since the ODE system (\ref{pde03}) is in general not analytically solvable, a numerical scheme is usually used to obtain solutions in the whole radial range compatible with the prescribed behavior near $r=0$ and close to $r\simeq \infty$.

We come back now to  the non-linear $\mathbb{CP}^1$-sigma model characterized at the beginning of this Section. The compatibility relation (\ref{potential}) between metric in the target space and potential fixes the scalar field self-interaction in the south chart to be:
\begin{equation}
U_S(|\phi^S|^2;\alpha^2) = \frac{1}{8} \Big[ \frac{4\rho^2 |\phi^S|^2}{\rho^2 + |\phi^S|^2} -\alpha^2 \Big]^2 \, \, .\label{potencialsur}
\end{equation}
If $\alpha \in (0,2\rho)$ a spontaneously broken symmetry scenario arises from the structure of the potential (\ref{potencialsur}). The family of degenerate vacua, which in the south chart coordinates reads
\begin{equation}
{\cal M}=\Big\{ v^S \in \mathbb{C}: |v^S| = \frac{\alpha \rho}{\sqrt{4\rho^2-\alpha^2}} \Big\} \, \, ,
\label{vacua}
\end{equation}
constitutes a parallel circle on the $\mathbb{S}^2$-sphere target manifold. The mass of the vector particle is the coefficient of the quadratic term $A_\mu A^\mu$ in the Lagrangian after choosing one point in the vacuum circle and expanding the scalar field near this point, $\phi^S(x^\mu)=\vert v^S\vert+\varphi_1^S(x^\mu)+i\varphi_2^S(x^\mu)$:
\begin{equation}
m_V^2=\frac{4\rho^4 \vert v^S\vert^2}{(\rho^2+\vert v^S\vert^2)^2}= \alpha^2\left(1-\frac{\alpha^2}{4\rho^2}\right) \, \, .\label{vecm}
\end{equation}
Simili modo, expanding the purely scalar field contribution to the Lagrangian in the unitary gauge where the spurious Goldstone boson is set to zero, $\varphi_2^S=0$,
\begin{eqnarray*}
&& {\cal L}^S(\vert v^s\vert, \varphi_1^S(x^\mu),\varphi_2^S(x^\mu)=0)=\frac{1}{2}\frac{4\rho^4}{(\rho^2+\vert v^S\vert^2)^2}\partial_\mu \varphi_1^S(x^\mu)\partial^\mu \varphi_1^S(x^\mu)-
\frac{1}{8}\left(\frac{4\rho^2(\vert v^S\vert+\varphi_1^S)^2}{\rho^2+(\vert v^S\vert+\varphi_1^S)^2}-\alpha^2\right)^2 \\ && \hspace{1cm}\simeq \frac{4\rho^4}{(\rho^2+\vert v^S\vert^2)^2}\Big[\partial_\mu \varphi_1^S(x^\mu)\partial^\mu \varphi_1^S(x^\mu)-\frac{1}{2}\frac{4\rho^2\vert v^S\vert^2}{(\rho^2+\vert v^s\vert^2)^2}\cdot \varphi_1^S(x^\mu)\varphi_1^S(x^\mu)\Big]+o((\varphi_1^S)^3)
\end{eqnarray*}
up to the quadratic order in the Higgs field $\phi_1^S(x^\mu)$ one finds that the scalar particle mass is also:
\begin{equation}
m_H^2=\frac{4\rho^4 \vert v^S\vert^2}{(\rho^2+\vert v^S\vert^2)^2}= \alpha^2\left(1-\frac{\alpha^2}{4\rho^2}\right) \, \, .\label{scam}
\end{equation}
At least in the south chart our model lives: 1) In a Higgs phase, all the particles have mass. 2) At the critical point between Type I
and Type II superconductivity, scalar and vector particles have the same mass.

Coordinates in the north chart are related to those in the south chart by means of the (orientation reversing) transition function: $\phi^N=\frac{\rho^2}{\phi^{*S}}$. In the norh chart the potential (\ref{potencialsur}) thus reads:
\begin{equation}
U_N(|\phi^N|^2;\alpha^2) = \frac{1}{8} \Big[ \frac{4\rho^2 |\phi^N|^2}{\rho^2 + |\phi^N|^2} - 4\rho^2+\alpha^2 \Big]^2 \, \, . \label{potencialnorte}
\end{equation}
The vacuum orbit seen in the north chart is the parallel circle:
\begin{equation}
\vert v^N\vert^2= \Big(\frac{4\rho^2}{\alpha^2}-1\Big)\rho^2 \label{vacuan} \, .
\end{equation}
Because $\vert v^N\vert^2=\rho^4\cdot \vert v^S\vert^{-2}$ one realizes that it is the same parallel as (\ref{vacua}) but seen from the south pole. Moreover, the masses of the scalar and vector particles in the north chart
\[
m_V^2=m_H^2= \frac{4\rho^4 \vert v^N\vert^2}{(\rho^2+\vert v^N\vert^2)^2}=\alpha^2 \Big(1-\frac{\alpha^2}{4\rho^2}\Big)
\]
are the same as in the south chart and the model is globally defined in target space.

The vacuum orbit parallel ${\cal M}$ splits the sphere into two spherical caps, each cap entirely belonging to either the south or the north chart. This fact is behind the existence of two species of self-dual vortices. Each species takes values only on one of these two caps sharing only the parallel ${\cal M}$ as common boundary. We recall also that stereographic coordinates describing the vortex scalar field $\Phi$ vanish at the vortex center. The two species of self-dual vortices in the gauged $\mathbb{CP}^1$-model are classified as follows:

\begin{enumerate}
\item \textit{The south vortex species:} If the $\mathbb{S}^2$-valued $\Phi$ field describing the vortex solution points downwards at the vortex center, $\Phi_3(0)=-1$, only the south chart enters the game. The first-order ODE system (\ref{pde03}) determining the south vortex species reads
\begin{equation}\frac{d f_n^S(r)}{d r} = \pm \frac{n}{r} f_n^S(r) [1-\beta_n^S(r)] \hspace{0.5cm},\hspace{0.5cm} \frac{n}{r} \frac{d \beta_n^S}{d r} = \frac{1}{2} \Big[ \alpha^2 - \frac{4\rho^2 (f_n^S(r))^2}{\rho^2 + (f_n^S(r))^2}] \Big]
\label{pde03sur}
\end{equation}
where the obvious notation $\phi^S(r,\theta)=f_n^S(r) e^{in\theta}$ has been used.

\item \textit{The north vortex species:} If the vortex configuration in $\Phi$-space valued at the origin points upwards, $\Phi_3(0)=1$, only the north chart plays a r$\hat{\rm o}$le. The self-duality ODE system solved by the north vortex species corresponds to
\begin{equation}
\frac{d f_n^N(r)}{d r} = \pm \frac{n}{r} f_n^N(r) [1-\beta_n^N(r)] \hspace{0.5cm},\hspace{0.5cm} \frac{n}{r} \frac{d \beta_n^N}{d r} = \frac{1}{2} \Big[ 4\rho^2 - \alpha^2 - \frac{4\rho^2 (f_n^N(r))^2}{\rho^2 + (f_n^N(r))^2}] \Big]
\label{pde03norte}
\end{equation}
derived from (\ref{pde03}) by taking the north chart fields and the appropriate function $G$.
\end{enumerate}

Although the ODE systems (\ref{pde03sur}) and (\ref{pde03norte}) characterize two different species of vortex solutions, they exhibit related features. Notice that the potentials describing the scalar field interactions in each chart are connected by the relation
\[
U_N(|\phi|^2;\alpha^2) = U_S(|\phi|^2;4\rho^2 - \alpha^2)
\]
where the unified notation has been again used. This identity implies that the functional form of the first-order ODE systems (\ref{pde03sur}) and (\ref{pde03norte}) is symmetric under the interchange of the parameters $\alpha^2$ and $4\rho^2-\alpha^2$; simply look at the right members of the ODE's on the right in (\ref{pde03sur}) and (\ref{pde03norte}). These two parameters are inversely related, the greater the first one is, the lesser the second one becomes since $\alpha^2\in (0,4\rho^2)$. The values of $\alpha^2$ and $4\rho^2-\alpha^2$ have a direct impact in the magnitudes of the derivatives of the function $f$ and $\beta$ as we can check from (\ref{pde03sur}) and (\ref{pde03norte}). The asymmetry between the values of $\alpha^2$ and $4\rho^2-\alpha^2$ causes the difference between the core sizes of the vortices belonging to different species. Recall that the vortex core is characterized by small values of the scalar field. In Figure 1, below we have displayed the field profiles of these two species of vortices with vorticity (or winding number) $n=3$ for $\alpha=1$ and $\rho=1$, tantamount to $4\rho^2-\alpha^2=3$.

In the first and second graphics in Figure 1 the scalar and vector field profiles of the $3$-vortex belonging to the south species are respectively depicted by means of arrow plots. The shadowed discs are centered at the zeroes of the scalar and vector fields.  In the third and fourth graphics analogous arrow plots for the north species of the $3$-vortex are shown. The components of the arrows specify the real and imaginary parts of the complex scalar fields. We observe in these figures that the winding number of the two species of $3$-vortex profiles is three: note that in any quadrant every arrow rotates a $\frac{3}{2}\pi$ angle. On these figures we have plotted density representations of the modulus of the scalar field. The darker the colour is the lesser the value of the modulus is in such a way that the shadowed region in the scalar field graphics represents the vortex core. Notice that for our choice of $\alpha$ and $\rho$ the south species vortices are thicker than their north species partners. The same routine has been followed for the representation of the vortex vector field profiles of the two species, see Figures 1-(2) and 1-(4).

\begin{figure}[h]
\centering
\includegraphics[height=4cm]{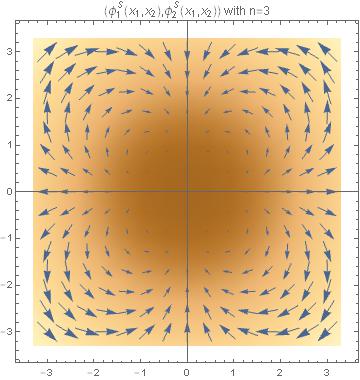} \includegraphics[height=4cm]{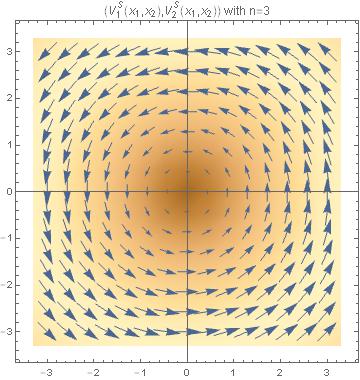}  \hspace{0.5cm} \includegraphics[height=4cm]{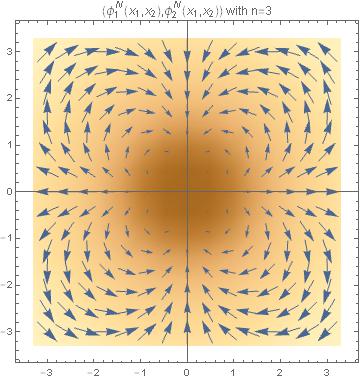}
\includegraphics[height=4cm]{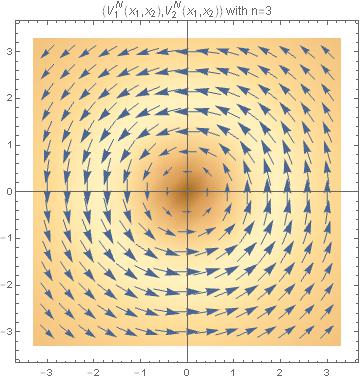}
\caption{Cross sections of the cylindrically symmetric self-dual 3-vortices in the south chart (first and second plots) and in the north chart (third and fourth plots). Graphics of the scalar field $\phi(\vec{x})$ (first and third figures) and the vector field $V(\vec{x})$ (second and fourth figures) profiles are depicted by means of Mathematica vector and density plots superimposed.}
\end{figure}

\section{Zero mode fluctuations around cylindrically symmetric BPS vortices of the two species}

We shall devote this Section to investigate the zero modes (eigenmodes with null eigenvalue) which exist in the spectrum of cylindrically symmetric vortex fluctuation operators of the two species. In \cite{Alonso} the existence has been shown of $2n$ linearly independent zero modes of fluctuation around BPS vortices living alternatively in one of the two charts by using the Callias-Bott-Weinberg index theorem, see \cite{Callias,Bott,Weinberg}. These eigenmodes are the key ingredient to describe the adiabatic motion of vortices in their moduli space, see e.g. \cite{Wifredo}, and therefore they are the cornerstone for studying the effective vortex dynamics in low energy scenarios.

\subsection{Analytical description of the zero modes of fluctuation of cylindrically symmetric BPS vortices}

In order to distinguish between the vortex solution and its fluctuations we shall denote the scalar and the vector field profiles corresponding to a self-dual vortex solution of vorticity $n$ as
\[
\psi(\vec{x};n) = \psi_1(\vec{x};n)+i \, \psi_2(\vec{x};n) \hspace{0.5cm},\hspace{0.5cm} V(\vec{x};n)=(V_1(\vec{x};n),V_2(\vec{x};n)) \, ,
\]
while the scalar and vector field fluctuations of the self-dual vortex solution will be denoted as $\varphi(\vec{x}) = \varphi_1(\vec{x}) + i\varphi_2(\vec{x})$ and $a(\vec{x})=(a_1(\vec{x}),a_2(\vec{x}))$ respectively. Fluctuations of this type built around the BPS vortex are zero modes of fluctuation if the perturbed fields
\[
\phi(\vec{x})=\psi(x_1,x_2;n)+\epsilon \varphi(\vec{x}) \hspace{0.6cm} \mbox{and} \hspace{0.6cm} A(\vec{x})=V(x_1,x_2;n)+ \epsilon a(\vec{x})
\]
are still solutions of the equations (\ref{pde01}) at first order in $\epsilon$. This requirement compels the fluctuation fields $a_j(\vec{x}), j=1,2$, and $\varphi_a(\vec{x}), a=1,2$ to comply with the first order partial differential equations
\clearpage
\begin{eqnarray}
-\partial_2 a_1+\partial_1 a_2 + g(|\psi|^2) (\psi_1 \varphi_1+\psi_2 \varphi_2) &=&0 \nonumber \\
\psi_1a_1-\psi_2 a_2 +(-\partial_2+V_1)\varphi_1 + (-\partial_1-V_2) \varphi_2 &=&0 \label{fluctu01} \\
\psi_2a_1+\psi_1 a_2 +(\partial_1+V_2)\varphi_1 + (-\partial_2+V_1) \varphi_2 &=&0 \nonumber
\end{eqnarray}
In order to discard pure gauge fluctuations we select the generalized background gauge in the fluctuation space
\begin{equation}
-\partial_1 a_1-\partial_2 a_2 + g(|\psi|^2) (\psi_1 \varphi_2-\psi_2 \varphi_1) =0 \label{backgroundgauge}\, \, .
\end{equation}
Assembling the field fluctuations in a real four column vector
\[
\xi^T(x_1,x_2)=\left( \, \, a_1(x_1,x_2) \, \, \, \,  a_2(x_1,x_2)\, \, \, \, \varphi_1(x_1,x_2) \, \, \, \, \varphi_2(x_1,x_2) \, \, \right)
\]
the first-order PDE system of the four equations (\ref{fluctu01})-(\ref{backgroundgauge}) may be re-written in the matrix form:
\begin{equation}
{\cal D}\xi(x_1,x_2)=\left( \begin{array}{cccc}
-\partial_2 & \partial_1 & g(|\psi|^2)\,\psi_1 & g(|\psi^2|)\,\psi_2 \\
-\partial_1 & -\partial_2 & -g(|\psi|^2)\,\psi_2 & g(|\psi^2|)\,\psi_1 \\
\psi_1 & -\psi_2 & -\partial_2+V_1 & -\partial_1-V_2 \\
\psi_2 & \psi_1 & \partial_1+V_2 & -\partial_2 +V_1
\end{array} \right) \left( \begin{array}{c} a_1 \\ a_2 \\ \varphi_1 \\ \varphi_2 \end{array} \right)=\left(\begin{array}{c} 0 \\ 0 \\ 0 \\ 0 \end{array}\right) \label{matrix} \, \, .
\end{equation}
Here, ${\cal D}$ denotes the first-order partial differential matrix deformation operator and the zero mode fluctuations $\xi(\vec{x})$ are required also to be square-integrable vectors of a Hilbert space ${\cal H}=\oplus L^2_a(\mathbb{R}^2)$ with $L^2(\mathbb{R}^2)$-norm
\begin{equation}
\|\xi(\vec{x})\|^2 =\int_{\mathbb{R}^2} d^2 x \Big[ (a_1(\vec{x}))^2 + (a_2(\vec{x}))^2  + g(|\psi|^2)(\varphi_1(\vec{x}))^2 + g(|\psi|^2)(\varphi_2(\vec{x}))^2 \Big]< +\infty \, \, . \label{norma00}
\end{equation}
We are interested in the zero mode fluctuations of cylindrically symmetric $n$-vortex solutions, and thus we plug the ansatz
\begin{eqnarray*}
\psi_1(\vec{x},n) =f_n(r)\cos(n\theta) \hspace{0.5cm} &,& \hspace{0.5cm} \psi_2(\vec{x},n) =f_n(r)\sin(n\theta) \\  V_1(\vec{x},n) =-\frac{n \beta_n(r)}{r} \sin \theta \hspace{0.5cm} &,& \hspace{0.5cm} V_2(\vec{x},n) =\frac{n \beta_n(r)}{r} \cos \theta
\end{eqnarray*}
into the PDE system (\ref{fluctu01})-(\ref{backgroundgauge}). The outcome is the equivalent PDE system:
\begin{eqnarray}
-\frac{1}{r} \frac{\partial a_r}{\partial \theta} + \frac{\partial a_\theta}{\partial r} + \frac{1}{r} a_\theta + f_n(r) \, g[f_n^2(r)] \cos(n\theta) \varphi_1 + f_n(r) \, g[f_n^2(r)] \sin(n\theta) \varphi_2&=&0 \nonumber \\
-\frac{\partial a_r}{\partial r} - \frac{1}{r} a_r - \frac{1}{r} \frac{\partial a_\theta}{\partial \theta} - f_n(r) \, g[f_n^2(r)] \sin(n\theta) \varphi_1 + f_n(r)\, g[f_n^2(r)] \cos(n\theta) \varphi_2&=&0 \label{fluctu02} \\
-\frac{1}{r} \frac{\partial \varphi_1}{\partial \theta} -\frac{\partial \varphi_2}{\partial r} - \frac{n \beta_n(r)}{r} \varphi_2 + f_n(r) a_r \cos(n\theta) -f_n(r) a_\theta \sin(n\theta)&=&0 \nonumber\\
-\frac{\partial \varphi_1}{\partial r} - \frac{n \beta_n(r)}{r} \varphi_1 + \frac{1}{r} \frac{\partial \varphi_2}{\partial \theta} - f_n(r) a_r \sin(n\theta) -f_n(r) a_\theta \cos(n\theta)&=&0 \nonumber
\end{eqnarray}
where we have used the polar representation for the vector fluctuation field: $a_1=a_r \cos \theta-a_\theta \sin\theta$ and $a_2=a_r\sin \theta+a_\theta \cos \theta$. In the system (\ref{fluctu02}) a discrete symmetry between the vectors in the kernel of the first-order operator ${\cal D}$ is hidden: if $\xi=(a_r,a_\theta, \varphi_1,\varphi_2)$ is a zero mode (a solution of the system (\ref{fluctu02})) then $\xi^\perp=(a_\theta,-a_r,\varphi_2,-\varphi_2)$ is another zero mode orthogonal with respect to $\xi$. This symmetry reduces the search for the $2n$ vortex zero modes of this field theory model to only $n$ of them. We now investigate the analytical properties of these eigenfunctions. Given the cylindrical symmetry of the vortex solutions, we propose an angular dependence of the zero mode wave unctions of the form
\begin{eqnarray}
&& a_r(r,\theta)=s_{nk}(r)\sin [(n-k)\theta] \hspace{1cm},\hspace{1cm} \varphi_1(r,\theta)=t_{nk}(r) \cos (k\theta) \nonumber \\
&& a_\theta(r,\theta)=s_{nk}(r) \cos[(n-k)\theta] \hspace{1cm},\hspace{1cm} \varphi_2(r,\theta)=t_{nk}(r) \sin (k\theta) \label{ansatz08}
\end{eqnarray}
where $k\in \mathbb{Z}$ in order to deal with single valued eigenfunctions: $\xi(r,\theta,n,k) = \xi(r,\theta+2\pi,n,k)$. Plugging the ansatz (\ref{ansatz08}) into the system of four first-order PDE's (\ref{fluctu02}) reduces the problem to solve the following system of two coupled first-order ODE's:
\begin{eqnarray}
\frac{d s_{nk}(r)}{d r} + \frac{1}{r} (k+1-n) s_{nk}(r) + f_n(r) \, g[f_n^2(r)]\,t_{nk}(r)&=&0 \nonumber \\
\frac{d t_{nk}(r)}{d r} + \frac{n \beta_n(r)-k}{r} t_{nk}(r) + f_n(r) s_{nk}(r) &=& 0 \label{pde05}
\end{eqnarray}
for the radial form factors $s_{nk}(r)$ and $t_{nk}(r)$ of the vortex zero mode fluctuations defined in (\ref{ansatz08}). Notice that the norm (\ref{norma00}) in the Hilbert space of normalizable fluctuations is simplified to \begin{equation}
\|\xi(\vec{x},n,k)\|^2 = 2\pi \int dr\, r\Big[ s_{nk}^2(r) + g(f^2_n(r))\, t_{nk}^2(r) \Big] \, \, .
\label{norma01}
\end{equation}
We can go further in our analytical calculations by solving for the radial function $t_{nk}(r)$ in the first equation of (\ref{pde05})
\begin{equation}
t_{nk}(r)=\frac{-1}{f_n(r) g[f_n^2(r)]} \Big[ \frac{d s_{nk}(r)}{d r} + \frac{1}{r}(k+1-n) s_{nk}(r) \Big] \label{funciont}
\end{equation}
and inserting the result in the second equation of (\ref{pde05}). This manoeuvre leads to the second-order linear differential equation for the radial form factor $s_{nk}(r)$:
\begin{eqnarray}
&& \hspace{-0.5cm} -r^2 g[f_n^2(r)] \frac{d^2 s_{nk}(r)}{d r^2} + r \Big[ g[f_n^2(r)] (-1+2n-2n\beta_n(r)) +2nf_n^2(r)(1-\beta_n(r))g'[f_n^2(r)]\Big] \frac{d s_{nk}(r)}{d r} +  \nonumber \\
&& \hspace{1cm} + \Big[g[f_n^2(r)](1+k-n)(1+k+n-2n\beta_n(r))+r^2f_n^2(r)(g[f_n^2(r)])^2 + \label{fluctu05} \\ &&\hspace{1.5cm} + 2n(1-\beta_n(r))(k+1-n) f_n^2(r) g'[f_n^2(r)]\Big] s_{nk}(r)=0 \nonumber \, \, .
\end{eqnarray}
In order to analyze the structure of zero mode vortex fluctuations we must study thus the properties of $L^2$-integrable solutions of (\ref{fluctu05}). We start this task by analyzing the behavior of the solutions of (\ref{fluctu05}) near the origin and subsequently we shall examine its asymptotic behavior far away from the vortex center:
\begin{itemize}
\item \textit{Regularity of the function $s_{nk}(r)$ at the origin.} The Frobenius method may be applied to the linear differential equations (\ref{fluctu05}) at the regular singular point $r=0$. We expand the function $s_{nk}(r)$ as a power series
    \begin{equation}
    s_{nk}(r)=r^s \sum_{j=0}^\infty c_j^{(n,k)} r^j =r^s h_{nk}(r) \hspace{0.8cm},\hspace{0.8cm} h_{nk}(r)= \sum_{j=0}^\infty c_j^{(n,k)} r^j \label{fluctu04}
    \end{equation}
where the power $s$ is chosen as the minimum value such that $c_0^{(n,k)}\neq 0$, i.e., $h_{nk}(r)$ is regular and does not vanish at the origin. The norm (\ref{norma01}) is now rewritten in terms of the power series $h_{nk}(r)$:
    \begin{equation}
    \|\xi(\vec{x},n,k)\|^2 = 2\pi \int dr \, r^{2s+1} \Big[ h_{nk}^2(r) + \frac{(h'_{nk}(r))^2}{g[f_n^2(r)]\,f_n^2(r)} \Big] \label{norma02}
    \end{equation}
by using the relation (\ref{funciont}) and (\ref{fluctu04}). Plugging the power series expansion (\ref{fluctu04}) into (\ref{fluctu05}) we obtain the following recurrence relation between the coefficients $c_j^{(n,k)}$:
\begin{eqnarray}
&&\sum_{j=0}^{2n+1} \Big[ (1+j+k-n+s)(1-j+k+n-2)g(0) c_j^{(n,k)} \Big] r^j + \nonumber \\ && + \sum_{j=2}^{2n+1} \Big[ -g(0)2n e_2 (-1+j+k-n+s)c_{j-2}^{(n,k)} \Big]r^j + \label{recurrence03} \\
&& \sum_{j=2n}^{2n+1} \Big[ g'(0)d_n^2 \Big( (-1+4n)(s+j-2n)+(1+k-n)(1+k+3n) \Big) c_{j-2n}^{(n,k)}  \Big] r^j + {\cal O}(r^{2n+2}) =0 \nonumber
\end{eqnarray}
where we have used the power series expansion (\ref{fseries}) and (\ref{bseries}) of the $n$-vortex radial form factors $f_n(r)$ and $\beta_n(r)$ together with the expansion of the metric factor evaluated in the vortex solution near its center:
\[
g[f_n^2(r)] = g(0)+g'(0) \,r^{2n} \, \Big(\sum_{\ell=0}^\infty d_{n+2\ell} \,\, r^{2\ell} \Big)^2
\]
Notice that in (\ref{recurrence03}) the recurrences are cut at order $2n+1$, we will see shortly that there is no need of taking into account more terms to ensure regularity at the origin and, henceforth, $L^2$-integrability, accounting for only the dominant terms near the vortex center.

For $j=0,1$ the recurrence (\ref{recurrence03}) is simply:
\begin{equation}
(1+j+k-n+s)(1-j+k+n-s)c_j^{(n,k)} =0 \label{recurrence04}
\end{equation}
Since from hypothesis $c_0^{(n,k)}\neq 0$, the indicial equation, $j=0$ in (\ref{recurrence04}) fixes the value of the characteristic exponents: $s=n-k-1$ and $s=n+k+1$. Both possibilities are equivalent: simply redefine $k$, $k \to -k-2$. Thus, we shall stick to the first option in the sequel. This choice of $s$ in equation (\ref{recurrence04}) for the index $j=1$ implies that necessarily $c_1^{(n,k)}=0$. Near the origin, the first summand in the integrand of (\ref{norma02}), recall that $s=n-k-1$, is therefore:
\[
r^{2(n-k)-1}h_{nk}^2(r) \simeq (c_0^{(n,k)}) r^{2(n-k)-1}+o(r^{2(n-k)+1}) \, \, .
\]
Poles at the origin in the integrand are skipped if
\begin{equation}
2(n-k)-1\geq 0 \, \, \, \Rightarrow \, \, \, k \leq n-1  \, \, , \label{degen}
\end{equation}
a condition which implies that the integer number $k$ is bounded by the vorticity $n$.

The two term recurrence relations for the next group of indices $j=3,\dots,2n-1$ and the characteristic exponent $s=n-k-1$ becomes:
\begin{equation}
j(2k+2-j) \, c_j^{(n,k)}=2n \, e_2 (j-2) \,c_{j-2}^{(n,k)}
\label{recurrence05}
\end{equation}
Starting from $c_1^{(n,k)}=0$ it is easily checked that (\ref{recurrence05}) implies $c_{2i+1}^{(n,k)}=0$  for all the odd indices $j=2i+1$ in the range $3<2i+1\leq 2n-1$. The recurrence (\ref{recurrence05}) for even indices, however, $j=2i$ reads:
\begin{equation}
i(k-i+1)c_{2i}^{(n,k)} =e_2(i-1)n c_{2i-2}^{(n,k)} \label{recurrence06} \, \, .
\end{equation}
Insertion of the values $i=1,2, 3,\dots,k$ in (\ref{recurrence06}) means that all the coefficients $c_2^{(n,k)}=c_4^{(n,k)}= \cdots =c_{2k}^{(n,k)}=0$ vanish. $c_2^{(n,k)}$ is zero because the factor $i(k-i+1)$ appearing in the left hand side of (\ref{recurrence06}) is non null, while $i-1$ present in the right-hand side is zero, for $i=1$. If $2\leq i \leq k$ a similar situation happens: all the right side members in (\ref{recurrence06}) are zero because the coefficients are zero but the left hand sides must be also zero, restricting the values of the coefficients up to $c_{2k}^{(n,k)}$ to be zero.  The first non-null coefficient after $c_0^{(n,k)}$ is $c_{2k+2}^{(n,k)}$ because $k-i+1=0$ in this case. The first two terms of the $s_{nk}(r)$-power series expansion near $r=0$ are thus:
\[
h_{nk}(r)=c_0^{(n,k)} + c_{2k+2}^{(n,k)}\, r^{2k+2} + O(r^{2k+3})
\]
where $c_0^{(n,k)}$ and $c_{2k+2}^{(n,k)}$ are arbitrary non-null constants. Together with the bound (\ref{degen}) this means that it is enough to identify the even coefficients up to $c_{2n+1}^{(n,k)}$ in order to describe the zero mode wave functions near the origin, a fact that justify the truncation assumed in the recurrence relations (\ref{recurrence03}). We finally pass to analyze the second summand in the integrand of (\ref{norma02}) near the origin:
\[
r^{2(n-k)-1} \frac{(h'_{nk}(r))^2}{g[f_n^2(r)]\,f_n^2(r)} = (c_{2k+2}^{(n,k)})^2 \frac{(2k+2)^2}{g(0)} r^{2k+1} + O(r^{2k+3}) \, \, ,
\]
seeing that it is regular at the origin if and only if $2k+1\geq 0$, i.e., $k\geq 0$.
Therefore regularity at the origin restricts the values of $k$ to the first $n$ natural numbers $k=0,1,2,\dots,n-1$ such that
there are at most $n$ zero modes, or rather $2n$, if the orthogonal zero modes to these null potential eigenfunctions are accounted for.

\item \textit{Asymptotic behavior of the function $s_{nk}(r)$.} For large values of $r$ the modulus of the scalar complex field tends to a constant value $f(r)\rightarrow f_\infty >0$ that belongs to the vacuum circle ${\cal M}$, whereas the radial profile of the vector field tends to one: $\beta(r)\rightarrow 1$. Bearing this asymptotic behavior in mind, we see that at large $r$ the ODE equation (\ref{fluctu05}) reduces to the modified Bessel differential equation:
    \[
    -r^2 \frac{d^2 s_{nk}(r)}{d r^2} - r \frac{d s_{nk}(r)}{d r} +  \Big[ (1+k-n)^2+r^2 \, f_\infty^2 \, g(f_\infty)  \Big] s_{nk}(r)=0 \, \, .
    \]
    The general solution of this second-order ODE is well known:
    \begin{eqnarray}
    s_{nk}(r) &\simeq& C_1 \, I_{1+k-n}\left[f_\infty\sqrt{g(f_\infty)} \, r\right]+ C_2 \, K_{1+k-n}\left[f_\infty\sqrt{g(f_\infty)} \,r\right] \nonumber \\ & \simeq & \overline{C}_1 \, \frac{1}{\sqrt{r}} \, e^{f_\infty\sqrt{g(f_\infty)} r} + \overline{C}_2 \, \frac{1}{\sqrt{r}} \, e^{-f_\infty\sqrt{g(f_\infty)} r} \label{asintoticog} \, \, ,
    \end{eqnarray}
    where $I_N[x]$ and $K_N[x]$ are modified Bessel functions respectively of the first and second kind. It is crystal clear that we must choose  $\overline{C}_1=0$ In formula (\ref{asintoticog}) in order to obtain zero mode eigenfunctions with an exponential decaying tail which satisfy the $L^2$-integrability condition.

\item \textit{Intermediate regime.} After describing analytically the eigenfunctions in the kernel of ${\cal D}$ near and far away from the vortex center, Sturm-Liouville guarantees the existence of a regular solution at the origin $r=0$ of the equation (\ref{fluctu05}) for every $k=0,1,\dots,n-1$ which has a decreasing exponential tail by simply tuning the values of the constants $c_0^{(n,k)}$ and $c_{2k+2}^{(n,k)}$ in order to obtain a solution with the adequate asymptotic behavior. In conclusion, there exists $n$ zero modes $\xi(\vec{x};k)$ of the generic form (\ref{ansatz08}) whose radial profiles $s_{nr}(r)$ and $t_{nr}(r)$ are solutions of the linear first-order ODE system (\ref{pde05}). Moreover, all these zero modes characterized by the wave number $k$ are linearly independent. Integration in the angular variable shows that these eigenfunctions are orthogonal:
\[
\int_0^{2\pi} \, d\theta \, \xi^T(r,\theta;k_1)\cdot \xi(r,\theta;k_2)=\delta_{k_1k_2}\cdot F^T(r;k_1) F(r;k_2)  \, \, .
\]
Together with their corresponding orthogonal partners $\xi^\perp(\vec{x};k)$, these whole set of $2n$ zero modes form a basis in the tangent space to the moduli space of BPS vortices at the cylindrically symmetric vortex carrying magnetic flux $n$ and centered at the origin.

Sturm-Liouville theory is enough to ensure the existence of these null eigenfunctions in the intermediate range between a neighborhood of the origin and other one close to the infinite point. Nevertheless, there is no way of analytically finding the vortex solutions at intermediate range, It is possible, however, to gather good information about the BPS vortex zero mode profiles by using numerical methods. In this sense it is better than to directly attack equation (\ref{fluctu05}) for $s_{nk}(r)$ simply to solve by numerical procedures the simpler equation in terms of the function $h_{nk}(r)$. Plugging
    \begin{equation}
    s_{nk}(r)=r^{n-k-1}h_{nk}(r) \label{gyh}
    \end{equation}
    in (\ref{fluctu05}) we end with the second-order linear ODE
    \begin{eqnarray}
    && \hspace{-0.5cm} -r g[f_n^2(r)] \frac{d^2h_{nk}(r)}{dr^2} + \Big[ g[f_n^2(r)](1+2k-2n\beta_n(r))+2 n f_n^2(r) g'[f_n^2(r)] (1-\beta_n(r)) \Big] \frac{dh_{nk}(r)}{dr} + \nonumber \\
    && \hspace{0.5cm} + r f_n^2(r) [g(f_n^2(r))]^2 h(r)=0 \label{fluctu09}
    \end{eqnarray}
    which will be our starting point to generate the zero mode fluctuation by means of the numerical scheme by some variant of a shooting procedure using the known solution near the origin as initial condition.
\end{itemize}

\subsection{Deformations of BPS cylindrically symmetric vortices of the two species by their zero mode fluctuations}

The perturbed fields up to first-order in the fluctuations
\[
\widetilde{\psi}(\vec{x};n,k) = \psi(\vec{x};n)+\epsilon \varphi(\vec{x},k)\hspace{0.5cm} \mbox{and} \hspace{0.5cm}\widetilde{V}(\vec{x};n,k)=V(\vec{x};n)+ \epsilon a(\vec{x},k)
\]
are deformed vortex solutions of the general BPS equation (\ref{pde01}). Recall that $\psi(\vec{x};n)$ and $V(\vec{x};n)$ stand respectively for the cylindrically symmetric self-dual vortex scalar and vector field profiles obtained through the ansatz (\ref{ansatz06}) and the solution of the BPS equations (\ref{pde03}). $\varphi(\vec{x},k)$ and $a(\vec{x},k)$ are the vortex zero mode fluctuations determined by using the ansatz (\ref{ansatz08}) to solve (\ref{funciont}), a procedure simplified by assuming (\ref{gyh}) to end with the  solution of (\ref{fluctu09}) as explained in the previous Section. The main r$\hat{\rm o}$le in the description of these perturbed solutions is played by the scalar field profile. In cylindrical coordinates adapted to the symmetry of the vortex flux lines the perturbed vortex complex field
profile reads:
\[
\widetilde{\psi}(\vec{x};n,k) = f_n(r) e^{in\theta}-\epsilon r^{n-k-1} \frac{h_{nk}'(r)}{g[f_n^2(r)] f_n(r)} e^{ik\theta} \, \, .
\]
Sufficiently close to the origin this profile is analytically known in terms of the integration constants $d_n$ and $c_{2k+2}^{(n,k)}$
plus the value of the metric at the origin $g(0)$:
\begin{equation}
\widetilde{\psi}(\vec{x};n,k) \approx r^k e^{ik\theta} \Big[ d_n r^{n-k} e^{i(n-k)\theta} - \epsilon \frac{(2k+2)c_{2k+2}^{(n,k)}}{g(0) d_n} \Big] \label{perturbed01}
\end{equation}
Because if
\[
r^k=0 \hspace{1cm} \mbox{and} \hspace{1cm} r^{n-k}=\epsilon \frac{(2k+2)c_{2k+2}^{(n,k)}}{g(0) d_n^2}e^{-i(n-k)\theta}
\]
the perturbed vortex scalar field profile (\ref{perturbed01}) exhibits one zero of multiplicity $k$ situated at the origin. We observe that the multiplicity of the zero at the origin decrease from $n$ to $k$ with respect to that of the unperturbed vortex scalar field.
$n-k$ new zeroes appear located at the vertices of a regular $k$-polygon:
\begin{eqnarray*}
r^{n-k} = \epsilon \frac{(2k+2)|c_{2k+2}^{(n,k)}|}{g(0) d_n^2} \hspace{0.5cm} && \mbox{and} \hspace{0.5cm} e^{i(n-k)\theta}={\rm sign}(c_{2k+2}^{(n,k)})=-1 \\ r=\Big( \epsilon \frac{(2k+2)|c_{2k+2}^{(n,k)}|}{g(0) d_n^2} \Big)^{\frac{1}{n-k}} \hspace{0.5cm} && \mbox{and} \hspace{0.5cm} \theta_j=\frac{2j+1}{n-k}\pi \, \, \, \, , \, \, \, \, \, j=0,1,2, \cdots, n-k-1  \, \, .
\end{eqnarray*}
slightly displaced from the single zero of multiplicity $n$ of the cylindrically symmetric vortex profiles placed at the vortex center.
In sum, under the $k=n-1$ zero mode perturbation one quantum of magnetic flux moves away from the origin along the $x$-axis while the remaining $n-1$ quanta stay at the origin. Under the next zero mode perturbation $k=n-2$ two quanta move respectively along the half-axis forming angles $\pi/2$ and $3 \pi/2$ with the $x_1$-axis, the other $n-2$ quanta staying at the origin. Under the generic $n-k$ zero mode  $n-k$ quanta of magnetic flux depart from the origin along the half-axis forming respectively angles $\theta_j$ with the $x_1$-axis, the other $k$ quanta remaining at the origin.

Perturbations of BPS cylindrically symmetric $n=3$-vortices in the $\mathbb{CP}^1$-sigma model undergoing zero mode fluctuations are illustrated in Figures 2 and 3 where the parameters $\rho=1$ and $\alpha =1$ are chosen. On BPS vortices with $n=3$ quanta of magnetic flux there exist six zero modes of fluctuation $\xi(\vec{x};k)$ and $\xi^\perp(\vec{x};k)$, characterized by the \lq\lq polarizations\rq\rq{} $k=0,1,2$. We recall that in the gauged $\mathbb{U}(1)$ $\mathbb{CP}^1$-model there are two species of BPS cylindrically symmetric vortices taking values respectively in the south and north charts of the target manifold. In Figure 2 the graphical information relative to the zero mode fluctuations $\xi^S(\vec{x},3,k)$ corresponding to the south species of BPS $3$-vortices is collected. In the first row the $k=2$ zero mode fluctuation $\xi^S(\vec{x},3,2)$ of the generic form (\ref{ansatz08}) is described. The scalar and vector field profiles $\varphi^S(\vec{x},3,2)$ and $a^S(\vec{x},3,2)$ of this null eigenmode $\xi^S(\vec{x},3,2)$ are respectively depicted in the first and the third graphics by means of a vectorial plot. In the second and fourth pictures of the same row it is plotted the way in which the scalar and vector fields of the circularly symmetric BPS 3-vortex configuration are deformed by the zero mode fluctuation $\xi^S(\vec{x},3,2)$.  One of the three single quanta of magnetic flux superimposed at the origin in the unperturbed solution is displaced along the $x_1$-axis while the remainder ones are untouched, see Figure 2 (first row, second plot). The same pattern is shown in the second row where the main features of the $k=2$ zero mode fluctuation $\xi^S(\vec{x},3,1)$ around a cylindrically symmetric BPS $3$-vortex belonging to the south chart are graphically described. In the first and third plots of this second row we see vectorial graphs of the scalar and vector fluctuations.
Perturbations where two quanta of magnetic flux are ejected from the vortex center are shown in the second and fourth graphs.
Finally, the third row in Figure 2 includes the plots corresponding to the $k=0$ zero modes $\xi^S(\vec{x},3,0)$. It is remarkable to notice that the three single vortices initially situated at the origin are expelled in the directions determined by the vertices of a equilateral triangle, see Figure 2 (third row, second plot).

In Figure 3 the same graphical information is collected and offered for the zero mode fluctuations $\xi^N(\vec{x},3,k)$ of the BPS cylindrically symmetric $3$-vortex belonging to the north chart species as well as the associated perturbed fields $\widetilde{\psi}^N(\vec{x},3,k)$ and $\widetilde{V}^N(\vec{x},3,k)$. The zero mode structure in this chart is almost identical to the structure shown in Figure 2 corresponding to the south species of vortices. The only important difference is the fact that the north species of vortices described in Figure 3 exhibits an smaller core than the south chart ones displayed in Figure 2, in agreement with the same discrepancy unveiled in Section 2 between the south and north vortex cores.

\begin{figure}[H]
\centering
\includegraphics[height=4cm]{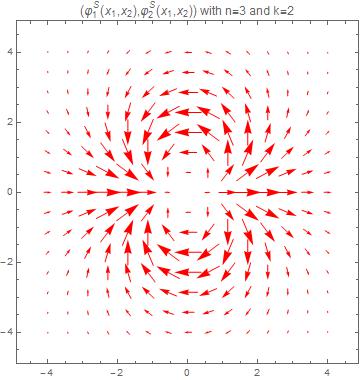} \includegraphics[height=4cm]{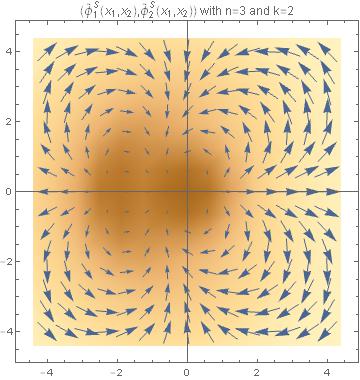}  \includegraphics[height=4cm]{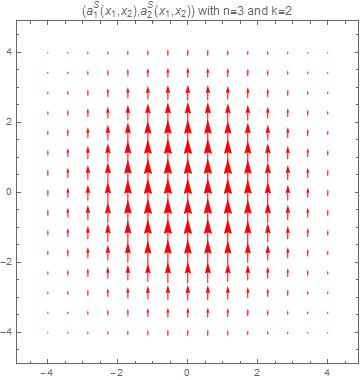}
\includegraphics[height=4cm]{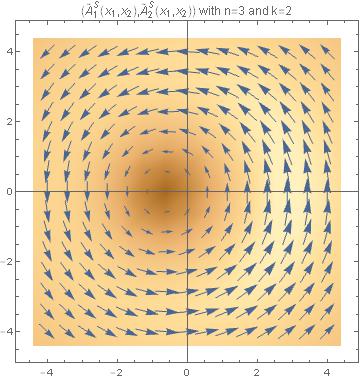} \\
\includegraphics[height=4cm]{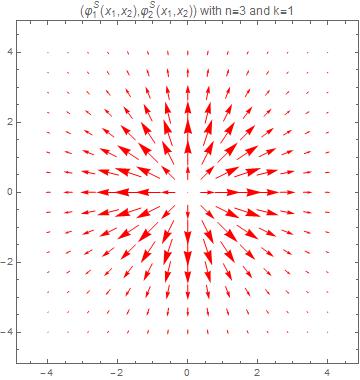} \includegraphics[height=4cm]{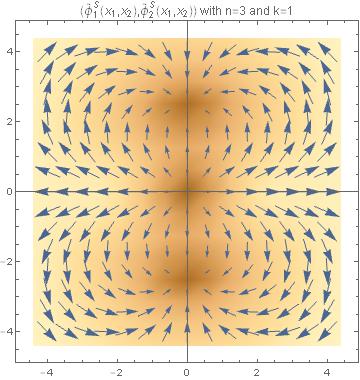}  \includegraphics[height=4cm]{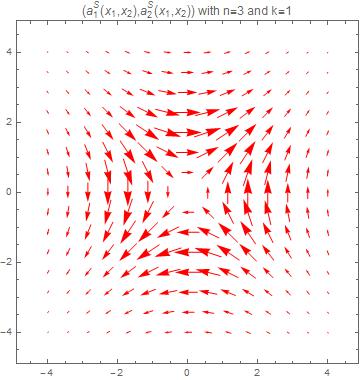}
\includegraphics[height=4cm]{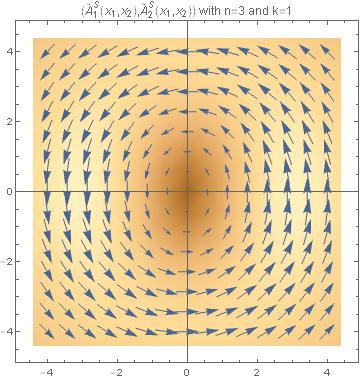} \\
\includegraphics[height=4cm]{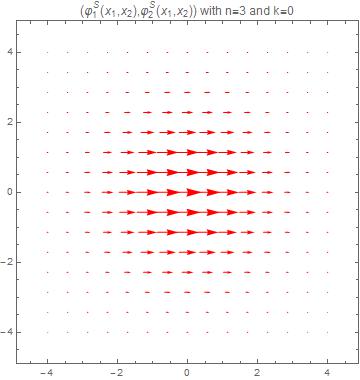} \includegraphics[height=4cm]{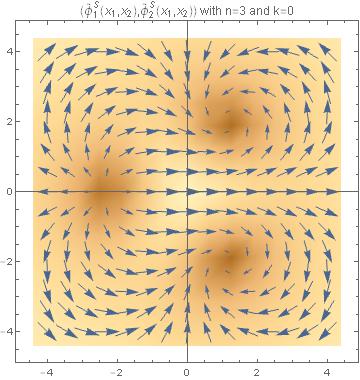}  \includegraphics[height=4cm]{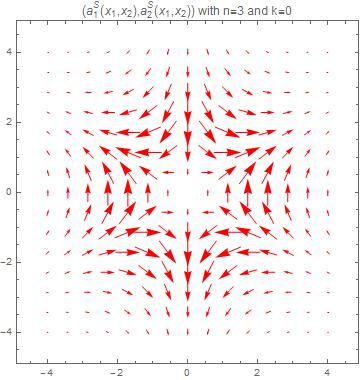}
\includegraphics[height=4cm]{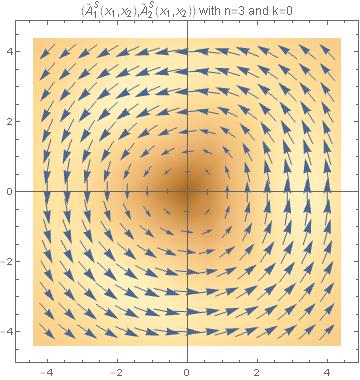}
\caption{Graphical representations of the scalar and vector components of the south class $3$-vortex zero mode fluctuations $\xi^S(\vec{x},3,k)$ (displayed in the first and third columns respectively) and the perturbed scalar and vector fields $\widetilde{\psi}^S(\vec{x},3,k)$ and  $\widetilde{V}^S(\vec{x},3,k)$ (displayed in the second and fourth columns respectively) for the values $k=2$ (first row), $k=1$ (second row) and $k=0$ (third row).}
\end{figure}

\begin{figure}[h]
\centering
\includegraphics[height=4cm]{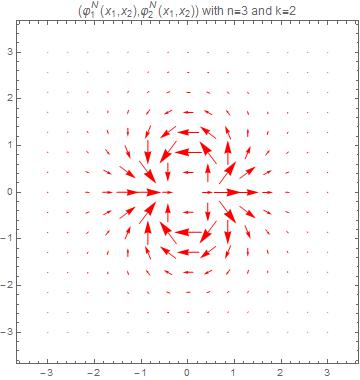} \includegraphics[height=4cm]{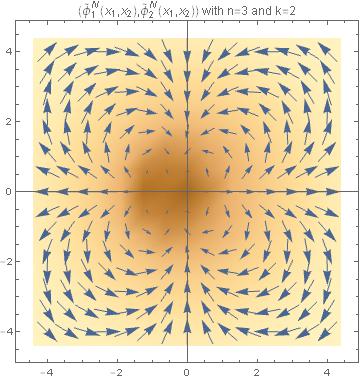}  \includegraphics[height=4cm]{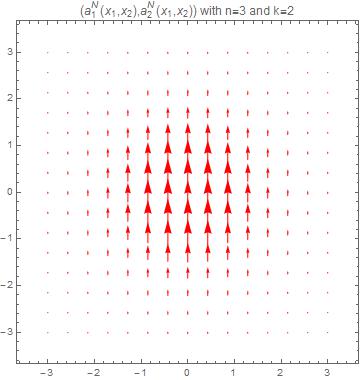}
\includegraphics[height=4cm]{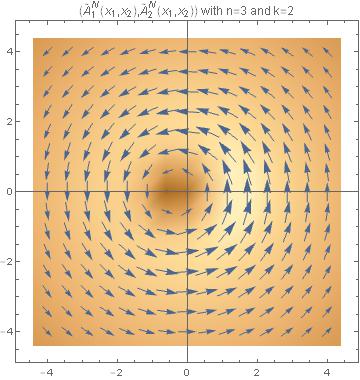} \\
\includegraphics[height=4cm]{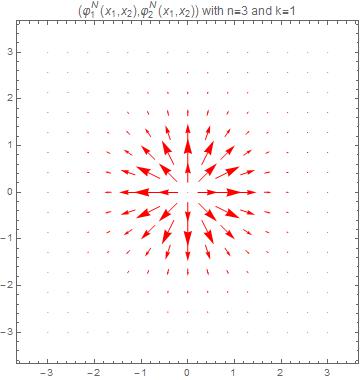} \includegraphics[height=4cm]{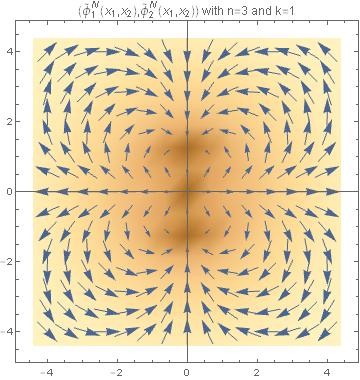}  \includegraphics[height=4cm]{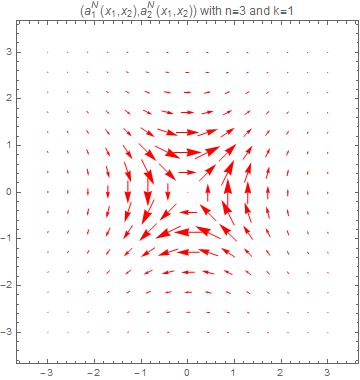}
\includegraphics[height=4cm]{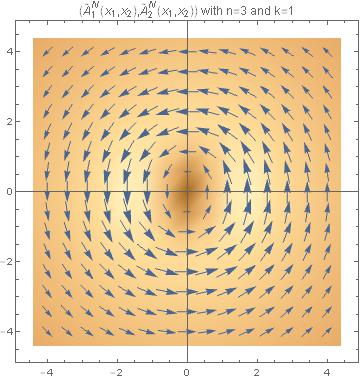} \\
\includegraphics[height=4cm]{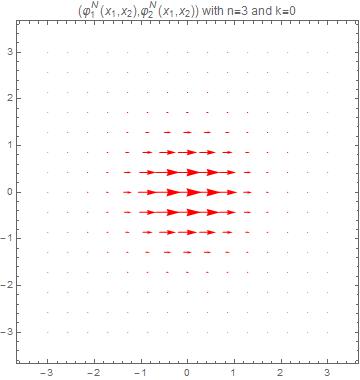} \includegraphics[height=4cm]{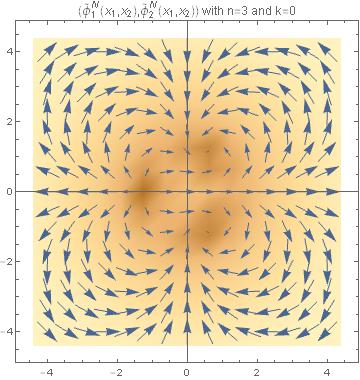}  \includegraphics[height=4cm]{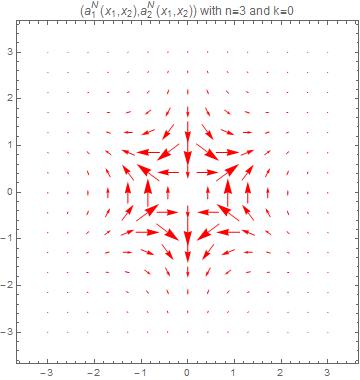}
\includegraphics[height=4cm]{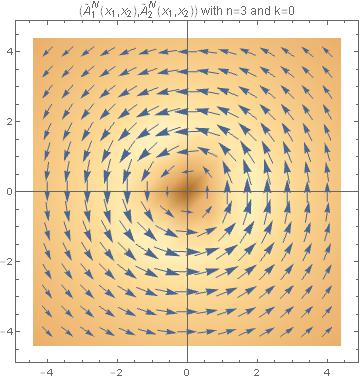}
\caption{Graphical representations of the scalar and vector components of the north class $3$-vortex zero mode fluctuations $\xi^N(\vec{x},3,k)$ (displayed in the first and third columns respectively) and the perturbed scalar and vector fields $\widetilde{\psi}^N(\vec{x},3,k)$ and  $\widetilde{V}^N(\vec{x},3,k)$ (displayed in the second and fourth columns respectively) for the values $k=2$ (first row), $k=1$ (second row) and $k=0$ (third row).}
\end{figure}

Despite the similarity between the structures described in Figures 2 and 3 the differences offer a relevant information about how the two species of vortices behave when they suffer a zero mode fluctuation. For the sake of comparison, we have employed the same magnitude of the perturbation parameter $\epsilon$ in the graphics representing perturbed fields in Figure 2 and 3. It is clear that the effect of the zero mode fluctuation in the north class of vortices is smaller than in the south class. In the context of adiabatic vortex dynamics we can claim that the thick (south class) self-dual $n$-vortices are split in its single constituents faster than the thin (north class) self-dual $n$-vortices by zero mode fluctuations, as we can see by comparing Figures 2 and 3.

\section{Outlook}

In this work we have thoroughly described the zero modes of fluctuation around cylindrically symmetric BPS vortices of the two species
existing in a gauge $\mathbb{U}(1)$ non-linear $\mathbb{CP}^1$ model that were discovered in Reference \cite{Alonso}. Besides their r$\hat{\rm o}$le in the scrutiny of low energy dynamics as achieved e.g. in \cite{Wifredo}, zero modes have an strong impact in the
evaluation of one-loop shifts to classical masses of BPS topological defects. It is conceivable to apply the method developed in
\cite{Alonso1} to the kinks discovered in Reference \cite{AlGuilLe} in order to improve the results obtained in \cite{AlGuilLe1}
about the one loop correction to the $\mathbb{S}^2$-kink masses. In a similar vein it seems highly plausible that the calculations
performed on the one-loop string tension shifts of the BPS vortices in the Abelian Higgs model in Reference \cite{AlGuilTor} may be
repeated successfully for the two species of BPS vortices in the $\mathbb{U}(1)$ gauge non-linear $\mathbb{CP}^1$ model profiting
from the results described in this paper. It is also tempting to extend this work to the $\mathbb{CP}^2$ and $\mathbb{CP}^N$
generalizations of the model treated here. In this case, besides Abelian vortices, semi-local topological defects also appear and
one might follow the developments in \cite{AlGuilTor1} in this non-linear context.

\end{document}